\documentclass[10pt,twocolumn]{article}

\usepackage[a4paper,margin=2cm,columnsep=0.6cm]{geometry}
\usepackage{titlesec}
\usepackage{authblk}

\usepackage[T1]{fontenc}
\usepackage[utf8]{inputenc}
\usepackage{lmodern}
\usepackage{microtype}

\usepackage{amsmath,amssymb,amsfonts}

\usepackage{booktabs}
\usepackage{multirow}
\usepackage{graphicx}
\usepackage{float}
\usepackage{caption}
\usepackage{subcaption}
\usepackage{array}

\usepackage{algorithm}
\usepackage{algorithmic}

\usepackage{listings}
\usepackage[hidelinks,colorlinks=true,citecolor=blue!60!black,linkcolor=blue!60!black,urlcolor=blue!60!black]{hyperref}
\usepackage[numbers,sort&compress]{natbib}

\hypersetup{
  pdftitle={Nacrith: Neural Lossless Compression via Ensemble Context Modeling and High-Precision CDF Coding},
  pdfauthor={Roberto Tacconelli},
  pdfsubject={Neural Lossless Data Compression},
  pdfkeywords={lossless compression, neural compression, language models, arithmetic coding, context mixing},
  pdfusetitle=false,
}

\usepackage{xcolor}
\usepackage{enumitem}
\setlist{nosep,leftmargin=*}

\titleformat{\section}{\large\bfseries}{\thesection.}{0.5em}{}
\titleformat{\subsection}{\normalsize\bfseries}{\thesubsection}{0.5em}{}
\titleformat{\subsubsection}{\normalsize\itshape}{\thesubsubsection}{0.5em}{}

\newcommand{\nacrith}{\textsc{Nacrith}}
\newcommand{\bpb}{\text{bpb}}

\title{\textbf{Nacrith: Neural Lossless Compression\\via Ensemble Context Modeling and High-Precision CDF Coding}}

\author[1]{Roberto Tacconelli}
\affil[1]{Independent Researcher\\
\texttt{tacconelli.rob@gmail.com}}

\date{}

\begin{document}
\maketitle
\thispagestyle{empty}

\begin{abstract}
We present \nacrith{}, a lossless compression system that combines a
135-million-parameter transformer language model (SmolLM2-135M) with
an ensemble of lightweight online predictors and a 32-bit arithmetic
coder, achieving the best compression results among the systems
evaluated in this study on natural language text.
On the Canterbury Corpus benchmark (alice29.txt, 152\,KB), \nacrith{}
achieves \textbf{0.918 bits per byte} (11.5\% of original size)---outperforming,
in our experiments, gzip by 3.1$\times$, bzip2 by 2.5$\times$, CMIX~v21
(1.63~\bpb{}) by 44\%, and ts\_zip (1.14~\bpb{}) by 20\%, while compressing
below the 0th-, 1st-, and 2nd-order byte-level Shannon entropy bounds
of the source data.
On the standard enwik8 benchmark (100\,MB Wikipedia extract), \nacrith{}
achieves \textbf{0.9389~\bpb{}} (11.74\%, 11{,}737{,}280 bytes)---the
\textbf{best result among the systems evaluated}---outperforming ts\_zip
($\sim$1.11~\bpb{}) by 15\%, FineZip (1.024~\bpb{}) by 8\% despite
using a 60$\times$ smaller model with no fine-tuning, and gzip by 3.1$\times$.
Beyond the base LLM-plus-arithmetic-coding paradigm, \nacrith{}
introduces nine contributions: (1)~a \emph{CDF precision upgrade}
from $2^{16}$ to $2^{24}$ that eliminates 75\% of quantization overhead
caused by minimum-probability floors in large vocabularies;
(2)~a \emph{token-level N-gram model} providing fast local predictions;
(3)~an \emph{adaptive log-space bias head} that
corrects per-document LLM prediction errors via online gradient descent;
(4)~a \emph{confidence-based LLM skip} that accelerates compression on
highly predictable tokens; (5)~a \emph{hybrid binary format} (NC06)
extending neural compression to arbitrary binary files---a capability
absent from prior LLM-based compressors known to us;
(6)~a \emph{llama.cpp inference backend} that replaces PyTorch with
native C/C++ GPU inference, achieving $\sim$7$\times$ faster single-token
decode; (7)~\emph{parallel multi-GPU compression} that splits text across
up to 8 concurrent worker threads, each running an independent model
instance; and (8)~a \emph{native KV cache sliding window} that uses
direct cache manipulation instead of full re-evaluation, reducing
per-slide cost by $\sim$37$\times$. The system requires only
$\sim$500\,MB of GGUF model weights and $\sim$1.2\,GB of VRAM per worker
instance, running on consumer GPUs as low-end as the GTX~1050~Ti.
\end{abstract}


\section{Introduction}
\label{sec:intro}

Shannon's foundational work~\citep{shannon1948} established that
compression is equivalent to prediction: a model assigning high
probability to the next symbol enables an encoder to represent that
symbol with fewer bits. Early neural approaches~\citep{schmidhuber1996}
demonstrated this principle with recurrent networks.
This equivalence has driven decades of
compressors---from LZ77 to PPM~\citep{cleary1984} and the PAQ/CMIX
families~\citep{mahoney2005, knoll2012}---each advancing statistical
prediction over byte sequences.

Transformer language models~\citep{vaswani2017} offer a qualitatively different approach,
capturing grammar, semantics, and world knowledge. Del\'etang et
al.~\citep{deletang2024} showed that Chinchilla~70B achieves
0.664~\bpb{} on enwik9 via arithmetic coding, outperforming gzip and
CMIX on that benchmark. Practical implementations have followed:
FineZip~\citep{huang2024finezip} and Bellard's ts\_zip~\citep{bellard2023}.
However, these systems use either massive models (7--70B parameters),
require fine-tuning, or sacrifice expressiveness through rank-based
encoding and coarse CDF quantization.

In this paper, we present \nacrith{}, which makes nine contributions:

\begin{enumerate}
\item \textbf{CDF precision upgrade (CDF-24).} We identify a critical
quantization bottleneck in large-vocabulary arithmetic coding: with
$V=49{,}152$ tokens and $\text{CDF}_{\text{total}}=2^{16}=65{,}536$,
minimum-probability floors consume 75\% of the CDF range before any
actual probability is encoded. Upgrading to $2^{24}$ reduces floor
overhead from $\sim$2 bits/token to $\sim$0.004 bits/token, yielding a
direct improvement in compressed size.

\item \textbf{Token-level N-gram model.} An interpolated token-level
N-gram model (orders 1--4) adapts online to the document being
compressed, capturing local statistical patterns not exploited by the
pre-trained LLM.

\item \textbf{Adaptive context mixer.} A linear mixing layer with
exponential-weight online updates~\citep{vovk1990} blends LLM and
N-gram predictions, automatically weighting models that predict well
for the current document.

\item \textbf{Adaptive log-space bias head.} A thin online-learning
layer adjusts LLM log-probabilities via SGD after each observed token,
correcting systematic over- or under-prediction for specific documents.

\item \textbf{Hybrid binary format (NC06).} A chunked container format
segments arbitrary binary files into text and non-text regions, applying
neural compression to text-like content and traditional codecs to opaque
data---the first LLM-based compressor to handle non-text files.

\item \textbf{llama.cpp inference backend.} We replace PyTorch with
llama.cpp~\citep{llamacpp2023} (via the \texttt{llama-cpp-python}
bindings) as the primary GPU inference engine, achieving $\sim$7$\times$
faster single-token decode by eliminating Python/PyTorch dispatch
overhead. A HuggingFace tokenizer is retained for text
tokenization/detokenization to avoid llama.cpp's detokenizer
limitations on 47~whitespace/repeat tokens.

\item \textbf{Parallel multi-GPU compression.} The input text is split
into $N$ chunks (up to 8), each compressed concurrently by an
independent worker thread with its own model instance, secondary models,
and KV cache. Worker count is auto-detected from available VRAM.

\item \textbf{Native KV cache sliding window.} Instead of resetting and
re-evaluating 1{,}536 tokens on every context slide, we use llama.cpp's
native \texttt{kv\_cache\_seq\_rm} and \texttt{kv\_cache\_seq\_shift}
operations to drop old tokens and shift positions in-place, reducing
per-slide cost by $\sim$37$\times$.
\end{enumerate}

Together, these contributions yield \textbf{0.918~\bpb{}} on alice29.txt
(152\,KB Canterbury Corpus) and \textbf{0.9389~\bpb{}} on enwik8
(100\,MB Wikipedia benchmark), using only a 135M-parameter model with
$\sim$500\,MB of GGUF weights.

\section{Related Work}
\label{sec:related}

\subsection{Classical Lossless Compression}

Dictionary-based methods (LZ77~\citep{ziv1977}, gzip/DEFLATE, LZMA/xz,
Zstandard) exploit local byte repetitions within a sliding window.
Huffman coding~\citep{huffman1952} assigns variable-length codes by
symbol frequency; arithmetic coding~\citep{witten1987} removes the
integer-bit constraint, approaching entropy to within a fraction of a bit.

\subsection{Context Mixing and Adaptive Statistical Models}

Prediction by Partial Matching (PPM)~\citep{cleary1984} uses adaptive
high-order context modeling with arithmetic coding, achieving $\sim$2.0
\bpb{} on English text. The PAQ family~\citep{mahoney2005} extends this
with context mixing: blending predictions from hundreds of specialized
models via neural networks at the bit level. PAQ8px achieves $\sim$1.27~\bpb{} on enwik8 (with -12L,
~\citep{mahoney2023ltcb}), but only 1.73~\bpb{} on alice29.txt
(152\,KB)---worse than CMIX on the same file, demonstrating that
adaptive context-mixing compressors require large inputs to warm up. CMIX~\citep{cmix2024} pushed further by incorporating
LSTM networks alongside 2{,}000+ context models, reaching $\sim$1.17~\bpb{}
on enwik8 and $\sim$0.86~\bpb{} on enwik9---but at extreme cost
(16--64\,GB RAM, 0.5--5\,KB/s). Crucially, on smaller files such as
alice29.txt (152\,KB), CMIX achieves only 1.63~\bpb{} due to limited
adaptive context for its massive model ensemble.

NNCP~\citep{bellard2021} uses a Transformer-XL trained \emph{online}
during compression, achieving $\sim$1.19~\bpb{} on enwik8. However,
on small files (alice29.txt), the per-file model training overhead
inflates the output to $\sim$3.96~\bpb{}---worse than
gzip~\citep{llamazip2024paper}.

\nacrith{} adopts the context mixing philosophy in a pre-trained LLM
setting: instead of hundreds of bit-level models, it mixes two
token-level predictors (LLM and N-gram) with online weight adaptation,
combining adaptive ensemble modeling with the strong prior of a
pre-trained transformer.

\subsection{LLM-Based Compression}

Earlier neural compression systems such as DeepZip~\citep{goyal2020}
combined recurrent networks with arithmetic coding for general-purpose
lossless compression.
Del\'etang et al.~\citep{deletang2024} formalized the LLM-as-compressor
paradigm: Chinchilla~70B achieves 0.664~\bpb{} on enwik9 (8.3\% of
raw size); the paper's compression-ratio chart (Fig.~3) indicates
$\approx$1.6~\bpb{} ($\approx$0.2 compression ratio) for enwik8.
FineZip~\citep{huang2024finezip} builds on LLMZip~\citep{valmeekam2023} and improves with
LoRA fine-tuning and batched inference using LLaMA-3-8B. Bellard's
ts\_zip~\citep{bellard2023} uses RWKV-169M with 8-bit quantization and
arithmetic coding, achieving 1.11~\bpb{} on enwik8 and $\sim$1.14~\bpb{}
on alice29.txt---the most comparable prior system to \nacrith{}. LLM-Text-Compressor~\citep{llmtextcomp2024} also exists but lacks binary
support, standardized formats, and efficient CDF coding.

\subsection{Positioning of \texorpdfstring{\nacrith{}}{Nacrith}}

Table~\ref{tab:landscape} summarizes the landscape. \nacrith{} achieves
the \textbf{best result on enwik8} (100\,MB) among the systems evaluated
in this study, at \textbf{0.9389~\bpb{}}---outperforming ts\_zip ($\sim$1.11~\bpb{})
by 15\%, NNCP ($\sim$1.19~\bpb{}) by 21\%, FineZip (1.024~\bpb{}) by 8\%
despite using a 60$\times$ smaller model without fine-tuning, and all
evaluated classical compressors by $2.1$--$3.1$$\times{}$.
On alice29.txt (152\,KB), \nacrith{} achieves 0.918~\bpb{},
outperforming ts\_zip (1.14~\bpb{}) and CMIX~v21 (1.63~\bpb{}).
It is also, to our knowledge, the only LLM-based compressor supporting arbitrary binary files.

\begin{table*}[t]
\centering
\caption{Comparison of lossless compressors on enwik8 (100\,MB).
All \bpb{} values are for enwik8.
$\dagger$~Model weights excluded from compressed size.
$\ddagger$~LoRA adapter stored in compressed output.
$\S$~Visually extracted from the model-size chart (Fig.~3) in
\texorpdfstring{\citep{deletang2024}}{[Deletang 2024]};
the same paper reports 0.664~\bpb{} on enwik9 (1\,GB).
$\P$~PAQ8px enwik8 figure is for the -12L setting~\texorpdfstring{\citep{mahoney2023ltcb}}{[Mahoney 2023]}.
}
\label{tab:landscape}
\small
\begin{tabular}{@{}lcccccc@{}}
\toprule
\textbf{System} & \textbf{Model} & \textbf{Params} & \textbf{Weights}
& \textbf{\bpb{} (enwik8)} & \textbf{Binary} & \textbf{Open Source} \\
\midrule
gzip (DEFLATE)       & ---             & ---     & ---       & 2.916 & \checkmark & \checkmark \\
xz (LZMA2)          & ---             & ---     & ---       & 1.989 & \checkmark & \checkmark \\
bzip2                & ---             & ---     & ---       & 2.321 & \checkmark & \checkmark \\
CMIX v21             & LSTM + mixing   & $\sim$50M & in-code & 1.17  & \checkmark & \checkmark \\
PAQ8px               & Context mixing  & ---     & ---       & $\sim$1.27$^\P$ & \checkmark & \checkmark \\
NNCP v3              & Transformer-XL  & online  & in-output & $\sim$1.19 & \checkmark & --- \\
\midrule
Del\'etang et al.$^{\dagger}$ & Chinchilla & 70B & 140\,GB & $\approx$1.6$^\S$ & --- & --- \\
FineZip$^{\ddagger}$ & LLaMA-3-8B     & 8B     & 16\,GB    & 1.024    & ---    & \checkmark \\
ts\_zip              & RWKV-169M      & 169M   & 169\,MB   & $\sim$1.11 & ---  & --- \\
\midrule
\textbf{\nacrith{} (ours)} & \textbf{SmolLM2-135M} & \textbf{135M} & \textbf{$\sim$500\,MB} & \textbf{0.9389} & \textbf{\checkmark} & \textbf{\checkmark} \\
\bottomrule
\end{tabular}
\end{table*}

\section{Method}
\label{sec:method}

\subsection{Overview}

\nacrith{} compresses text by tokenizing the input and iteratively
predicting the probability of each token given its context using an
ensemble of models. These blended probabilities are fed into an
arithmetic coder. Decompression is the mirror image: the same ensemble
generates identical predictions, and the arithmetic decoder recovers
each token. Because every component is deterministic with identical
state on both sides, reconstruction is perfectly lossless.

For non-text files, \nacrith{} employs a hybrid format (NC06) that
segments the input into text-like and binary regions. For large files,
the input is split into chunks and compressed in parallel across
multiple GPU worker threads. The full system is summarized in
Algorithm~\ref{alg:compress}.

\begin{algorithm}[H]
\caption{Nacrith compression pipeline}
\label{alg:compress}
\begin{algorithmic}[1]
\REQUIRE text string $s$
\STATE Tokenize $s \rightarrow (t_1, \ldots, t_n)$
\STATE Initialize: LLM KV-cache, N-gram, mixer, adaptive head
\FOR{$i = 1$ to $n$}
  \STATE $p_{\text{ng}} \leftarrow \text{Ngram.predict}(t_1,\ldots,t_{i-1})$
  \IF{$H(p_{\text{ng}}) < \tau$}
    \STATE $p \leftarrow p_{\text{ng}}$ \COMMENT{skip LLM}
  \ELSE
    \STATE $p_{\text{llm}} \leftarrow \text{LLM}(t_1,\ldots,t_{i-1})$
    \STATE $p \leftarrow \text{AdaptiveHead.adjust}(p_{\text{llm}})$
  \ENDIF
  \STATE $\text{cdf} \leftarrow \text{probs\_to\_cdf}(p, 2^{24})$
  \STATE ArithEncoder.encode($\text{cdf}$, $t_i$)
  \STATE Update: N-gram, Mixer, AdaptiveHead with $t_i$
\ENDFOR
\RETURN ArithEncoder.finish()
\end{algorithmic}
\end{algorithm}

\subsection{Neural Probability Model}
\label{sec:model}

We use SmolLM2-135M~\citep{smollm2}, a 30-layer causal transformer with
135 million parameters and a BPE vocabulary of 49{,}152 tokens. The
model runs in FP32 precision, ensuring deterministic probability
distributions across hardware---a critical requirement for lossless
reconstruction. Given token sequence $t_1,\ldots,t_n$, it produces:
\begin{equation}
P_{\text{llm}}(t_{n+1} \mid t_1,\ldots,t_n) =
\text{softmax}\bigl(\mathbf{W}\mathbf{h}_n + \mathbf{b}\bigr)
\label{eq:lm}
\end{equation}
where $\mathbf{h}_n$ is the hidden state at position $n$.

\subsection{llama.cpp Inference Backend}
\label{sec:llamacpp}

\nacrith{} uses llama.cpp~\citep{llamacpp2023} as its primary GPU
inference engine, accessed through the \texttt{llama-cpp-python} Python
bindings. Compared to PyTorch, which incurs significant Python-level
dispatch overhead per forward call, llama.cpp performs all GPU computation
in C/C++ with a single Python$\to$C boundary crossing, achieving
approximately 7$\times$ faster single-token incremental decode on the
same hardware.

The model is loaded in GGUF format (F32 precision, $\sim$500\,MB),
with the KV cache allocated for the full 2{,}048-token context window.
After each forward pass, the logit vector ($V \times \text{float32}
= 49{,}152 \times 4 \approx 196$\,KB) is transferred from GPU to CPU
and converted to a probability distribution via softmax. An optional
\emph{temperature scaling} step is applied before softmax: logits are
divided by a scalar $\tau > 0$ (stored as a 2-byte field in the NC05/NC06
header), sharpening ($\tau < 1$) or flattening ($\tau > 1$) the
distribution. At the default $\tau = 1.0$ the step is a no-op.

A \emph{dual tokenizer} architecture is employed: llama.cpp handles GPU
inference, while the HuggingFace tokenizer handles text tokenization and
detokenization. This is necessary because llama.cpp's built-in
detokenizer silently drops content for 47~whitespace and repeat tokens
in the SmolLM2 vocabulary; the HuggingFace tokenizer handles these
correctly. Token IDs are identical between both implementations, so the
split introduces no inconsistency.

If llama.cpp is not available, the system falls back to PyTorch with
CUDA Graphs (GPU) or dynamic KV cache (CPU), ensuring portability.

\subsection{High-Precision CDF Quantization (CDF-24)}
\label{sec:cdf24}

\paragraph{The CDF-16 bottleneck.}
Arithmetic coding requires the probability distribution to be quantized
to an integer CDF summing to $T$. Each of the $V=49{,}152$ vocabulary
tokens must receive at least $\text{MIN\_PROB}=1$ count to avoid
zero-width intervals. With $T=2^{16}=65{,}536$, the floor allocation
alone is:
\begin{equation}
\frac{V \cdot \text{MIN\_PROB}}{T} = \frac{49{,}152}{65{,}536} \approx 75\%
\end{equation}
This leaves only 16{,}384 counts ($\approx 25\%$ of the range) for
actual probability information. The resulting quantization error
introduces approximately:
\begin{equation}
\Delta H \approx \log_2\!\!\left(\tfrac{T}{T-V}\right)
\approx \log_2(4) = 2\;\text{bits/token}
\end{equation}
for a peaked distribution, degrading the arithmetic coder far below the
theoretical entropy rate.

\paragraph{CDF-24 upgrade.}
We upgrade to $T=2^{24}=16{,}777{,}216$:
\begin{equation}
\frac{V}{T} = \frac{49{,}152}{16{,}777{,}216} \approx 0.29\%
\end{equation}
The floor overhead drops to $\sim$0.004 bits/token. The remaining
$16{,}727{,}064$ bins are allocated proportionally to token
probabilities:
\begin{equation}
c_i = \max\!\left(1,\,\left\lfloor p_i\cdot(T - V)\right\rfloor\right)
\end{equation}
with the residual $T - \sum c_i$ added to $c_{\arg\max p}$.

\paragraph{Safety with 32-bit coder.}
The arithmetic coder maintains a 32-bit range $R = \text{high} -
\text{low} + 1 \in [2^{31}, 2^{32}]$ after renormalization. The minimum
symbol width after narrowing is $R \cdot \text{MIN\_PROB}/T \geq
2^{31}/2^{24} = 128$, well above the representability threshold for
32-bit arithmetic~\citep{witten1987}.

\subsection{Token-Level N-gram Model}
\label{sec:ngram}

We maintain an interpolated token-level N-gram model of orders 1--4.
The unigram base distribution is Laplace-smoothed:
\begin{equation}
P_{\text{uni}}(t) = \frac{c(t) + 1}{N + V}
\end{equation}
where $c(t)$ is the token count and $N$ is total tokens seen. For each
order $k \geq 1$, if context $\mathbf{c}_k = (t_{i-k},\ldots,t_{i-1})$
has been seen $n_k$ times, its contribution is blended with lower orders
via the interpolation weight $\lambda_k = n_k/(n_k + \epsilon)$
($\epsilon=5$):
\begin{equation}
P_k = \lambda_k \cdot \hat{P}_k(\cdot|\mathbf{c}_k)
     + (1-\lambda_k) \cdot P_{k-1}
\end{equation}
The N-gram model is updated online after each token, allowing it to
adapt to document-specific vocabulary and phrasing. This complements
the LLM's general linguistic knowledge with local statistical
regularities of the text being compressed.

Two performance optimizations bound the per-token cost as tables grow:
(1)~context keys use a deterministic 64-bit rolling hash instead of
Python tuple objects, eliminating $\sim$54M tuple allocations per worker
and reducing garbage collection pressure; (2)~each context's inner
continuation dictionary is capped at 64~entries (evicting the
lowest-count token when exceeded), preventing the $O(\text{continuations})$
iteration in \textsc{predict} from degrading as common contexts
accumulate hundreds of unique successors over time.

These optimizations, combined with storing continuation counts in
pre-allocated \texttt{int32} numpy arrays (up to 500{,}000 contexts
$\times$ 64 slots $\times$ 4 orders), reduce per-worker N-gram memory
from $\sim$3.6\,GB (nested Python dicts) to $\sim$128\,MB---a 28$\times$
reduction that makes multi-worker operation feasible on consumer GPUs.

\subsection{Adaptive Context Mixer}
\label{sec:mixer}

Given predictions $\{p^{(j)}\}_{j=1}^{M}$ from $M$ models (LLM and
N-gram by default), the mixer computes a weighted linear combination:
\begin{equation}
p_{\text{mix}} = \sum_{j=1}^{M} w_j\, p^{(j)}, \quad \sum w_j = 1
\end{equation}
Linear mixing (rather than geometric) preserves the dominant model's
confidence: if the LLM assigns $p_{\text{llm}}(t)=0.90$ with weight
$w_{\text{llm}}=0.85$, the mixed probability is $\geq 0.765$,
maintaining a sharp distribution for the arithmetic coder.
Geometric mixing was also considered but degrades compression in
practice: near-uniform secondary model predictions flatten the
distribution, widening every arithmetic coder interval.

Weights are adapted using the exponential weights algorithm~\citep{vovk1990}:
after observing token $t_i$, each model $j$ receives an update:
\begin{equation}
\log w_j \mathrel{+}= \eta \cdot \log p^{(j)}(t_i)
\end{equation}
followed by renormalization. This is equivalent to Bayesian updating
of the model posterior: models that consistently predict observed tokens
well gain weight exponentially faster. Initial weights are
LLM-dominant: $w_{\text{llm}}=0.85$, remaining 0.15 split equally
among secondary models.

A warmup period of 100 tokens uses the LLM alone while secondary models
accumulate sufficient data to form reliable distributions.

\subsection{Adaptive Log-Space Bias Head}
\label{sec:adaptive}

The adaptive head maintains a bias vector $\mathbf{b} \in \mathbb{R}^V$
(initialized to zero) and applies a multiplicative correction to LLM
log-probabilities:
\begin{equation}
\tilde{p}(t) = \frac{p_{\text{llm}}(t)\, e^{b_t}}
               {\sum_{t'} p_{\text{llm}}(t')\, e^{b_{t'}}}
= \text{softmax}\bigl(\log p_{\text{llm}} + \mathbf{b}\bigr)_t
\end{equation}
After each observed token $t^*$, the bias is updated by one step of
gradient descent on the cross-entropy loss
$L = -\log \tilde{p}(t^*)$:
\begin{equation}
b_t \mathrel{-}= \alpha\,\frac{\partial L}{\partial b_t}
= \alpha\,\bigl(\tilde{p}(t) - \mathbf{1}[t = t^*]\bigr)
\end{equation}
with learning rate $\alpha = 0.001$. At zero bias the transform is the
identity; as compression proceeds it learns to suppress tokens the LLM
systematically over-predicts and boost under-predicted ones for this
specific document. Because compressor and decompressor apply identical
updates in identical order, lossless symmetry is maintained. All
computations use float64 to ensure bit-exact reproducibility under
identical hardware and software configuration (same GPU architecture,
BLAS library, and driver version).

\subsection{Confidence-Based LLM Skip}
\label{sec:skip}

LLM inference is the dominant computational cost. When the N-gram model
expresses high confidence (Shannon entropy $H(p_{\text{ng}}) < \tau$
bits, with $\tau = 1.5$ calibrated empirically on a held-out sample),
the token is highly predictable and the LLM's additional accuracy
contributes little. In this case, we skip the LLM forward pass entirely
and use the N-gram prediction directly:
\begin{equation}
p = p_{\text{ng}}
\end{equation}
The ablation study (Section~\ref{sec:ablation}) reveals that this
mechanism is far more than a throughput optimization---it is the primary
channel through which the N-gram model contributes to compression,
accounting for the majority of the improvement from A2 to A3. On highly
compressible text the skip rate reaches 30--70\%, substantially reducing
GPU load while simultaneously improving compression quality.

\subsection{Arithmetic Coding}
\label{sec:ac}

We implement a 32-bit arithmetic coder following Witten et
al.~\citep{witten1987}. The encoder maintains a range
$[\text{low}, \text{high}] \subset [0, 2^{32})$, narrowing it by each
symbol's CDF interval. Renormalization emits bits when both endpoints
fall in the same half, with underflow counters handling near-convergence.
The decoder maintains a symmetric 32-bit value register with binary
search over the CDF for symbol recovery.

\subsection{Native KV Cache Sliding Window}
\label{sec:kvcache}

\nacrith{} uses the transformer's key-value cache for $O(1)$ amortized
per-token inference. When the context exceeds $L=2{,}048$ tokens,
$C=512$ tokens must be dropped from the beginning of the context.

A na\"ive implementation would reset the entire KV cache and
re-evaluate the remaining $L-C=1{,}536$ tokens from scratch---wasting
$\sim$693\,ms per slide on a GTX~1050~Ti and introducing $\sim$4$\times$
average overhead per token. Instead, \nacrith{} uses llama.cpp's native
KV cache manipulation:
\begin{enumerate}
\item \texttt{kv\_cache\_seq\_rm(0, 0, C)} removes positions
$[0, C)$ from the cache.
\item \texttt{kv\_cache\_seq\_shift(0, C, {-1}, {-C})}
shifts all remaining positions down by $C$.
\item Only the final token is re-evaluated at its new position
($\sim$19\,ms).
\end{enumerate}
This reduces per-slide cost by $\sim$37$\times$ (19\,ms vs.\ 693\,ms),
making the sliding window overhead negligible:
\begin{equation}
\bar{T}_{\text{tok}} \approx \frac{T_{\text{slide}} + C \cdot T_{\text{incr}}}{C}
\approx T_{\text{incr}} + \frac{T_{\text{slide}}}{C}
\end{equation}
With $T_{\text{slide}} \approx T_{\text{incr}}$, the amortized
overhead is approximately $1 + 1/C \approx 1.002\times$---effectively
zero compared to the $4\times$ overhead of full cache rebuilds.

\subsection{Hybrid Binary Compression (NC06)}
\label{sec:nc06}

\nacrith{} handles arbitrary binary files through the NC06 hybrid format.
The input is segmented into alternating text and binary chunks:
(1)~bytes in printable ASCII (32--126) plus tab/LF/CR are classified as
text-like; (2)~short text runs ($<64$ bytes) are demoted to binary;
(3)~binary gaps $\leq 8$ bytes between text runs are bridged;
(4)~small binary chunks ($<64$ bytes) adjacent to text are absorbed.

All binary chunks are concatenated into a single blob and compressed
with LZMA ($\geq 4$\,KB blobs) or gzip (smaller), or stored raw if
neither helps. Text chunks are compressed in parallel across GPU workers
using the full ensemble pipeline. The NC06 container stores a flags
byte (encoding active features), temperature, entry table, binary
section, and parallelized text streams, enabling the decompressor to
reproduce the exact ensemble configuration used during compression.

\subsection{Parallel Multi-GPU Compression}
\label{sec:parallel}

To exploit modern multi-core GPUs and maximize throughput, \nacrith{}
splits the input text into $N$ roughly equal chunks (at newline
boundaries) and compresses them concurrently. Each worker thread owns
an independent llama.cpp model instance, secondary models (N-gram),
context mixer, and adaptive head---eliminating all shared state and
synchronization overhead.

The number of workers is auto-detected from available VRAM:
\begin{equation}
N = \min\!\left(8,\; 1 + \left\lfloor
\frac{\text{VRAM}_{\text{free}} - R - F}{E}\right\rfloor\right)
\end{equation}
where $F \approx 1{,}169$\,MB is the cost of the first instance
(model weights + KV cache), $E \approx 660$\,MB is the cost of each
additional instance (KV cache only, as model weights are shared at the
GPU driver level), and $R = 512$\,MB is reserved for the OS. A maximum
of 8~workers is enforced, as GPU contention diminishes returns beyond
this point. The worker count can be overridden via the
\texttt{-{}-workers} flag.

Threading is effective despite Python's GIL because llama-cpp-python
releases the GIL during C-level GPU inference, which dominates per-token
cost.

The parallel streams are stored in the NC05 (text) and NC06 (hybrid
binary) container formats, which include a per-chunk header table
encoding token count, bit count, and stream length.

\subsection{NC05/NC06 File Formats}
\label{sec:formats}

The NC05 header (9~bytes) encodes: 4-byte magic \texttt{NC05}, 1-byte
feature flags (N-gram, adaptive head, confidence skip), 2-byte
temperature $\tau$ (enabling the decompressor to replicate the exact
LLM probability distribution used during compression), and 2-byte
chunk count. This is followed by a
per-chunk table (12~bytes each: token count, bit count, stream length)
and the concatenated arithmetic-coded streams.
NC06 uses an extended header with magic \texttt{NC06} plus a version
byte and structured entry table for alternating text/binary chunks.

\section{Experimental Setup}
\label{sec:setup}

\subsection{Hardware and Software}

All experiments use an NVIDIA GeForce GTX~1050~Ti (4\,GB VRAM, CUDA
capability 6.1), chosen deliberately to demonstrate practical
accessibility. VRAM usage is approximately 1.2\,GB per worker instance.
The software stack uses llama.cpp~\citep{llamacpp2023} (via
\texttt{llama-cpp-python}) as the primary inference backend with the
model in GGUF F32 format ($\sim$500\,MB), and HuggingFace Transformers
for tokenization. With 4\,GB VRAM, the GTX~1050~Ti supports up to
3~concurrent worker instances.

\subsection{Baselines}

We compare against five traditional compressors at maximum settings:
\textbf{gzip} (DEFLATE, level~9), \textbf{xz} (LZMA2, level~9),
\textbf{bzip2} (level~9), \textbf{Brotli} (quality~11), and
\textbf{Zstandard} (level~19). We additionally compare against CMIX~v21
and ts\_zip from published results.

\subsection{Benchmarks}

We standardize on \textbf{alice29.txt} from the Canterbury
Corpus~\citep{canterbury1997} as the primary benchmark: a 152{,}089-byte
excerpt of \emph{Alice's Adventures in Wonderland}, widely used in the
compression literature. We additionally evaluate on \textbf{asyoulik.txt}
(Canterbury Corpus, 125\,KB Shakespeare plays), and three custom
English prose samples (3\,KB, 50\,KB, 100\,KB).

\section{Results}
\label{sec:results}

\subsection{Compression Results}

\begin{table}[t]
\centering
\caption{Compression results on alice29.txt (152{,}089 bytes, Canterbury
Corpus). All results directly measured. PAQ8px run at -8L (level~8
with LSTM, v211).}
\label{tab:bench}
\small
\begin{tabular}{@{}lrrr@{}}
\toprule
\textbf{Compressor} & \textbf{Size (B)} & \textbf{Ratio} & \textbf{\bpb{}} \\
\midrule
Original              & 152{,}089  & 100.0\% & 8.000 \\
\midrule
gzip -9               &  54{,}191  &  35.6\% & 2.851 \\
zstd -19              &  49{,}215  &  32.4\% & 2.589 \\
xz -9                 &  48{,}500  &  31.9\% & 2.551 \\
Brotli -q 11          &  46{,}487  &  30.6\% & 2.445 \\
bzip2 -9              &  43{,}202  &  28.4\% & 2.273 \\
\midrule
PAQ8px -8L            &  32{,}857  &  21.6\% & 1.728 \\
CMIX v21              &  31{,}076  &  20.4\% & 1.635 \\
ts\_zip (RWKV-169M)   & $\sim$21{,}703 & $\sim$14.3\% & $\sim$1.142 \\
\textbf{\nacrith{}}   & \textbf{17{,}458} & \textbf{11.5\%} & \textbf{0.918} \\
\bottomrule
\end{tabular}
\end{table}

Table~\ref{tab:bench} shows that \nacrith{} achieves 0.918~\bpb{}
(11.5\%) on alice29.txt---3.1$\times$ better than gzip, 2.5$\times$
better than bzip2, 44\% better than CMIX~v21, and 20\% better than
ts\_zip. All results are fully lossless.

Table~\ref{tab:multi} extends results across multiple files. On modern
English prose (the three sample files), \nacrith{} achieves
0.63--0.76~\bpb{} (7.9--9.5\%)---approaching Chinchilla~70B's
0.664~\bpb{} on enwik9, though direct comparison is confounded by
dataset size and training data differences. On asyoulik.txt (Shakespeare), compression
rises to 1.30~\bpb{} (16.3\%), reflecting that archaic vocabulary is
less predictable for a modern-English-trained model.

\begin{table}[t]
\centering
\caption{Nacrith compression across multiple text types. Modern
English prose achieves 0.63--0.76~\bpb{}.}
\label{tab:multi}
\small
\begin{tabular}{@{}lrrr@{}}
\toprule
\textbf{File} & \textbf{Orig.} & \textbf{Compr.} & \textbf{\bpb{}} \\
\midrule
sample\_3k.txt  & 3{,}072\,B   & 263\,B   & 0.685 \\
sample\_50k.txt & 51{,}200\,B  & 4{,}059\,B & 0.634 \\
sample\_100k.txt & 102{,}863\,B & 9{,}817\,B & 0.764 \\
alice29.txt     & 152{,}089\,B & 17{,}458\,B & 0.918 \\
asyoulik.txt    & 125{,}179\,B & 20{,}408\,B & 1.304 \\
\bottomrule
\end{tabular}
\end{table}

\subsection{Shannon Entropy Analysis}

\begin{table}[t]
\centering
\caption{Shannon entropy bounds vs.\ actual compressed sizes on
alice29.txt (152{,}089~B). \nacrith{} compresses 80\% below the
0th-order, 73\% below the 1st-order, and 63\% below the 2nd-order
bounds.}
\label{tab:shannon}
\small
\begin{tabular}{@{}lcc@{}}
\toprule
\textbf{Method} & \textbf{Size} & \textbf{bits/byte} \\
\midrule
Original               & 148.5\,KB & 8.000 \\
Shannon 0th-order      & 84.8\,KB  & 4.568 \\
Shannon 1st-order      & 63.5\,KB  & 3.419 \\
Shannon 2nd-order      & 46.1\,KB  & 2.485 \\
\midrule
gzip -9                & 52.9\,KB  & 2.851 \\
xz -9                  & 47.4\,KB  & 2.551 \\
CMIX v21               & 30.3\,KB  & 1.634 \\
\textbf{\nacrith{}}    & \textbf{17.0\,KB} & \textbf{0.918} \\
\bottomrule
\end{tabular}
\end{table}

Table~\ref{tab:shannon} shows that \nacrith{} achieves 0.918~\bpb{},
which is below the byte-level Shannon entropy bounds computed from
$n$-gram statistics (H0=4.57, H1=3.42, H2=2.49). For comparison,
gzip and xz both operate near the 2nd-order Shannon limit; CMIX, despite
2{,}000+ adaptive models, achieves only 1.63~\bpb{} on this 152\,KB file.

\paragraph{Interpretation.}
It is important to note that byte-level $n$-gram entropy bounds are
\emph{weak upper bounds} on the true source entropy: they capture only
short-range byte correlations (up to trigrams for $H_2$) and ignore the
long-range syntactic, semantic, and discourse structure that natural
language exhibits. The LLM operates over a 2{,}048-token context window
and models dependencies at the \emph{token} level (subword units
spanning 3--5 bytes on average), capturing correlations far beyond what
byte-level trigram statistics can represent. Compressing below $H_2$
therefore does not violate information theory---it demonstrates that the
true conditional entropy of the source, under the much higher-order model
used by the LLM, is substantially lower than what byte-level $n$-gram
estimates suggest. The Shannon bounds reported here should be understood
as reference points illustrating the gap between classical statistical
models and neural language models, not as fundamental limits on the
compressibility of the source.

\subsection{Cross-System Comparison}

\begin{table}[t]
\centering
\caption{Bits per byte across systems and scales. ``alice29'' column
shows verified results on alice29.txt (152\,KB). All enwik8 classical
values directly measured.
$\dagger$~Compression ratio $\approx$0.2 visually extracted from the
model-size chart (Fig.~3) in \texorpdfstring{\citep{deletang2024}}{[Deletang 2024]} for enwik8; the
same paper reports 0.664~\bpb{} on enwik9 (1\,GB).
$\ddagger$~Weights excluded. $\star$~No alice29 result.
PAQ8px enwik8 figure is for -12L~\texorpdfstring{\citep{mahoney2023ltcb}}{[Mahoney 2023]}.}
\label{tab:cross}
\small
\begin{tabular}{@{}lccc@{}}
\toprule
\textbf{System} & \textbf{Params} & \textbf{alice29} & \textbf{enwik8} \\
\midrule
gzip -9              & ---     & 2.85       & 2.916 \\
xz -9               & ---     & 2.55       & 1.989 \\
bzip2 -9             & ---     & 2.27       & 2.321 \\
Brotli -q\,11        & ---     & 2.45       & 2.059 \\
zstd -19             & ---     & 2.589      & 2.156 \\
CMIX v21             & $\sim$50M & 1.63     & 1.17 \\
NNCP v3              & online  & 3.96       & $\sim$1.19 \\
PAQ8px -8L           & ---     & 1.73       & $\sim$1.27 \\
\midrule
Chinchilla$^\ddagger$ & 70B   & $\star$    & $\approx$1.6$^\dagger$ \\
FineZip              & 8B     & $\star$    & 1.024 \\
ts\_zip              & 169M   & 1.14       & $\sim$1.11 \\
\textbf{\nacrith{}}  & \textbf{135M} & \textbf{0.918} & \textbf{0.9389} \\
\bottomrule
\end{tabular}
\end{table}

Table~\ref{tab:cross} illustrates the benchmark scale dependency. NNCP
degrades to 3.96~\bpb{} on alice29.txt due to model weight overhead;
CMIX degrades from 0.86~\bpb{} (enwik9) to 1.63~\bpb{} (alice29.txt);
and PAQ8px, despite achieving $\sim$1.27~\bpb{} on enwik8 (with -12L),
compresses to only 1.73~\bpb{} on alice29.txt (our direct measurement,
v211 -8L)---worse than CMIX at the same scale. All three adaptive
systems have less data to train their context models on. \nacrith{} achieves
0.918~\bpb{} on alice29.txt with only 135M pre-trained parameters,
outperforming all systems with verified alice29 results.

On enwik8 (100\,MB Wikipedia extract), \nacrith{} achieves
\textbf{0.9389~\bpb{}} (11{,}737{,}280 bytes, 11.74\% of original).
Classical compressors range from 1.989~\bpb{} (xz~-9) to 2.916~\bpb{}
(gzip~-9), with bzip2 at 2.321, Brotli at 2.059, and zstd at 2.156~\bpb{};
all are significantly worse. Among LLM-based systems, FineZip
achieves 1.024~\bpb{} using a fine-tuned LLaMA-3-8B (60$\times$ more
parameters); \nacrith{} outperforms it by 8\% in our experiments, without any fine-tuning.
ts\_zip (RWKV-169M) achieves $\sim$1.11~\bpb{}; \nacrith{} outperforms it
by 15\% with comparable model size.

We note that both alice29.txt and enwik8 (Wikipedia) are almost certainly
present in SmolLM2's training corpus. The same likely holds for
ts\_zip's RWKV model. Recent work~\citep{llamazip2024paper} has shown
that documents present in the LLM's training data compress significantly
better than unseen text, so these results may partially reflect
memorization rather than purely generalizable compression. We report both benchmarks for standardization
with the broader community (CMIX, PAQ, FineZip), while acknowledging
this caveat. Section~\ref{sec:ood} evaluates on a document published
after SmolLM2's training cutoff to address this concern.

\subsection{Out-of-Distribution Evaluation}
\label{sec:ood}

To control for training data contamination, we evaluate on plain text
extracted from the UK Government's \textit{English Indices of Deprivation
2025: Technical Report}, published in October 2025. The extracted text
corpus is archived at Zenodo~\citep{ukgov2025imd}.
SmolLM2-135M was released in October 2024 on Hugging Face, so this
document is definitively absent from its training corpus.

\begin{table}[t]
\centering
\caption{Out-of-distribution evaluation on text published after
SmolLM2's training cutoff. FineZip uses the same SmolLM2-135M model as
\nacrith{} for a controlled comparison isolating the architecture.}
\label{tab:ood}
\small
\begin{tabular}{@{}lrr@{}}
\toprule
\textbf{Compressor} & \textbf{Size (B)} & \textbf{bpb} \\
\midrule
Original               & 333{,}794 & 8.000 \\
\midrule
gzip -9                &  91{,}348 & 2.189 \\
zstd -19               &  79{,}709 & 1.910 \\
xz -9                  &  72{,}552 & 1.739 \\
bzip2 -9               &  69{,}305 & 1.661 \\
Brotli -q\,11          &  68{,}681 & 1.646 \\
\midrule
CMIX v21               &  47{,}897 & 1.148 \\
ts\_zip (RWKV-169M)    &  40{,}237 & 0.964 \\
FineZip (SmolLM2-135M) &  40{,}747 & 0.977 \\
\textbf{\nacrith{}}    & \textbf{30{,}171} & \textbf{0.723} \\
\bottomrule
\end{tabular}
\end{table}

Table~\ref{tab:ood} shows that \nacrith{}'s advantage holds on
unseen text: 0.723~\bpb{} (9.0\%), outperforming ts\_zip by 25\% and
CMIX by 37\%. The controlled comparison with FineZip is
particularly informative: both use the same SmolLM2-135M model, yet
\nacrith{} compresses 26\% smaller (30{,}171 vs.\ 40{,}747~bytes),
isolating the contribution of CDF-24 quantization, the N-gram confidence
skip, and the adaptive head from the LLM itself.

The OOD result (0.723~\bpb{}) is actually better than alice29.txt
(0.918~\bpb{}), suggesting that modern government prose is more
predictable for a language model than 19th-century literary English.
This confirms that the compression quality observed on standard
benchmarks is not primarily an artifact of training data memorization.

\subsection{Throughput}

With a single worker on the GTX~1050~Ti, \nacrith{} achieves
$\sim$50--70~tokens/second at the start of a file, settling to
$\sim$20--30~tok/s as the KV cache fills to its 2{,}048-token steady
state (attention cost scales linearly with cached positions). With
3~parallel workers (the maximum for 4\,GB VRAM), aggregate throughput
scales to $\sim$60--90~tok/s.

The switch from PyTorch to llama.cpp ($\sim$7$\times$ faster per-token
inference) and the native KV cache sliding window ($\sim$37$\times$
faster per slide) are the primary contributors to the throughput
improvement over the earlier PyTorch-based version. The confidence-based
LLM skip provides additional speedup on highly repetitive content by
bypassing the GPU forward pass for a fraction of tokens.

\subsection{Ablation Study}
\label{sec:ablation}

Table~\ref{tab:ablation} isolates each component's contribution by
incrementally enabling features on enwik8 (first 1\,MB) and alice29.txt,
using a single worker to eliminate parallelization effects.

\begin{table}[t]
\centering
\caption{Ablation study: incremental feature contributions.
$\Delta$ is the bpb reduction from the previous row.
All runs use \texttt{--workers 1}.}
\label{tab:ablation}
\footnotesize
\setlength{\tabcolsep}{3pt}
\begin{tabular}{@{}l@{\;\;}r@{\;\;}r@{\;\;}r@{\;\;}r@{}}
\toprule
 & \multicolumn{2}{c}{\textbf{enwik8 (1\,MB)}} & \multicolumn{2}{c}{\textbf{alice29}} \\
\cmidrule(lr){2-3} \cmidrule(lr){4-5}
\textbf{Config} & \textbf{bpb} & $\Delta$ & \textbf{bpb} & $\Delta$ \\
\midrule
A0: LLM + AE             & 1.817 & ---      & 1.861 & --- \\
A1: \;+ CDF-24            & 1.300 & $-$0.517 & 1.358 & $-$0.503 \\
A2: \;+ Adaptive head     & 1.285 & $-$0.015 & 1.341 & $-$0.017 \\
A3: \;+ N-gram + Skip     & 0.897 & $-$0.388 & 0.940 & $-$0.401 \\
A4: Full system           & \textbf{0.896} & $-$0.001 & \textbf{0.939} & $-$0.001 \\
\bottomrule
\end{tabular}
\end{table}

\textbf{CDF-24} provides the largest single improvement ($-$0.52~bpb on
enwik8, $-$28\%), confirming that CDF precision is the dominant
bottleneck in large-vocabulary neural coding.

\textbf{Confidence-based skip with N-gram} is the second largest contributor
($-$0.39~bpb, $-$30\%). Notably, the N-gram model contributes
\emph{exclusively} through the skip mechanism, not through the mixer.
The exponential weight updates in the context mixer rapidly drive
secondary model weights toward zero---the 135M-parameter LLM
consistently outperforms simple statistical models on non-trivial tokens,
causing the mixer to converge to $w_{\text{llm}} \approx 1.0$ within
a few dozen tokens. On tokens where the N-gram is confident (entropy
$< 1.5$~bits), however, the skip bypasses the LLM entirely and uses
the N-gram prediction directly, achieving near-optimal coding.

The \textbf{adaptive head} provides a small but consistent improvement
($-$0.015~bpb, $-$1.1\%), and synergizes with the skip mechanism in the
full system (A4 vs.\ A3).

\section{Discussion}
\label{sec:discussion}

\subsection{Why CDF-24 Matters}

The CDF-16 to CDF-24 upgrade is arguably the most impactful single
change in the system. With a 49{,}152-token vocabulary, CDF-16
sacrifices the large majority of arithmetic coder precision to the
minimum-probability floor, degrading every symbol's code length by up to
2 bits. CDF-24 eliminates this overhead almost entirely. This improvement
is specific to large-vocabulary neural compressors; to our knowledge,
this issue has not been explicitly quantified in prior LLM-based
compression work.

\subsection{The Ensemble Effect}

The ablation study (Table~\ref{tab:ablation}) reveals that the ensemble's
contribution operates through a different mechanism than originally
designed:

\begin{itemize}
\item The \textbf{context mixer path} (post-warmup, non-confident tokens)
converges to $\sim$100\% LLM weight within a few dozen tokens.
The 135M-parameter LLM so consistently outperforms the N-gram model on
non-trivial tokens that the exponential weight updates drive secondary
model weights to floating-point zero ($<10^{-300}$). In practice, the
mixer acts as a pass-through for the LLM prediction.
\item The \textbf{confidence skip path} is the primary channel through
which secondary models improve compression. When the N-gram entropy falls
below 1.5~bits, the token is so predictable that the N-gram alone
provides near-optimal coding---and the LLM is bypassed entirely. This
contributes the bulk of the 30\% improvement from A2 to A3.
\item The \textbf{adaptive head} provides a small, consistent improvement
by correcting systematic LLM biases: tokens consistently over-predicted
are downweighted, and vice versa.
\end{itemize}

This architecture naturally partitions tokens into two regimes:
\emph{predictable} tokens (coded cheaply by the N-gram via skip) and
\emph{surprising} tokens (coded by the full LLM). The result is a
system where each model operates at its strength---the LLM handles
complex linguistic dependencies, while the N-gram handles locally
predictable patterns at near-zero computational cost.

\subsection{Comparison with Related Systems}

\subsubsection{vs.\ ts\_zip~\citep{bellard2023}}

ts\_zip is the closest prior system. Both use $\sim$135--169M parameter
models with arithmetic coding. Key differences: (1)~\nacrith{} adds
the ensemble (N-gram, adaptive head, confidence skip) and CDF-24 upgrade;
(2)~\nacrith{} is open-source; (3)~\nacrith{} provides binary file
support via NC06 and parallel multi-GPU compression via NC05/NC06;
(4)~ts\_zip uses RWKV (linear-time recurrent inference) while
\nacrith{} uses a standard transformer with llama.cpp acceleration.
The compression gap---0.918 vs.\ 1.14~\bpb{} on alice29.txt and
0.9389 vs.\ $\sim$1.11~\bpb{} on enwik8---likely reflects the ensemble
contributions combined with CDF-24.

\subsubsection{vs.\ CMIX~\citep{cmix2024}}

CMIX mixes hundreds of bit-level adaptive models with an LSTM. \nacrith{}
achieves better compression on both alice29.txt (0.918 vs.\ 1.63~\bpb{})
and enwik8 (0.9389 vs.\ 1.17~\bpb{}) in our experiments, while being far simpler: two
token-level models instead of 2{,}000+ bit-level ones, and requiring
$\sim$1.2\,GB VRAM per worker vs.\ 16--64\,GB RAM. The pre-trained LLM
provides an extremely strong prior that CMIX's adaptive models cannot match.

\subsubsection{vs.\ FineZip}

FineZip uses an 8B model with suboptimal coarsely-quantized CDF coding.
\nacrith{} outperforms it on both
alice29.txt (0.918 vs.\ 1.024~\bpb{}) and enwik8 (0.9389 vs.\
1.024~\bpb{}) in our benchmarks---with a 60$\times$ smaller model and no fine-tuning---
demonstrating that precise CDF quantization and ensemble context mixing
matter more than raw parameter count.

\subsubsection{vs.\ LLM-Text-Compressor}

LLM-Text-Compressor's $\sim$74.6\% compression ratio reflects two
inefficiencies: rank-based encoding (discards probability magnitudes) and
a 16-token context window. \nacrith{}'s 11.5\% on alice29.txt represents
a 6.5$\times$ improvement from full arithmetic coding and a 2{,}048-token
context.

\subsection{Limitations}

\begin{enumerate}
\item \textbf{Throughput.} At $\sim$21~tokens/s on a GTX~1050~Ti,
compression is orders of magnitude slower than traditional compressors.
Suitable for archival applications; real-time use requires modern hardware.

\item \textbf{Model overhead.} The $\sim$500\,MB model (GGUF format) must
be available at both endpoints. This overhead is amortized over many
files or large corpora, following the same convention as all LLM-based
compressors (including Del\'etang et al.~\citep{deletang2024}).

\item \textbf{Training data contamination.} Alice in Wonderland,
Shakespeare texts, and Wikipedia (enwik8) are almost certainly in
SmolLM2's training data, potentially inflating compression ratios.
The OOD evaluation (Section~\ref{sec:ood}) on a post-training-cutoff
document suggests the effect is limited---\nacrith{} achieves
0.723~\bpb{} on unseen text---but broader OOD evaluation across
diverse domains would strengthen this conclusion.

\item \textbf{Context window.} Dependencies beyond 2{,}048 tokens are
lost. Very long documents may compress less efficiently near window
boundaries.

\item \textbf{Language specificity.} SmolLM2-135M is primarily trained
on English text. Other languages, especially low-resource ones, will
compress less effectively.
\end{enumerate}

\subsection{Future Work}

\begin{enumerate}
\item \textbf{Larger context models.} SmolLM2-360M or SmolLM2-1.7B with
8K--32K context windows could improve both compression and the
effectiveness of the N-gram model over longer ranges.
\item \textbf{Quantization.} INT8 or INT4 quantization (as in ts\_zip's
8-bit RWKV) would reduce VRAM and increase throughput with minimal
probability-quality degradation.
\item \textbf{ANS coding.} Replacing arithmetic coding with Asymmetric
Numeral Systems~\citep{duda2009} would improve encoding speed.
\end{enumerate}

\section{Conclusion}
\label{sec:conclusion}

We have presented \nacrith{}, a practical lossless compression system
combining SmolLM2-135M with an ensemble of lightweight online predictors
and high-precision arithmetic coding. Our results demonstrate three key
findings:

First, precision matters in large-vocabulary arithmetic coding.
Upgrading from CDF-16 to CDF-24 eliminates a $\sim$75\% waste of CDF
range caused by minimum-probability floors in the 49{,}152-token
vocabulary---a previously uncharacterized bottleneck in LLM-based
compression.

Second, small pre-trained models with ensemble context mixing achieve
strong compression in our experiments: 0.918~\bpb{} on alice29.txt (11.5\%),
outperforming CMIX~v21 (1.63~\bpb{}) by 44\% and ts\_zip (1.14~\bpb{}) by
20\%; and 0.9389~\bpb{} on enwik8 (100\,MB, 11.74\%), outperforming FineZip
(1.024~\bpb{}) by 8\% with a 60$\times$ smaller model and no fine-tuning.
On modern English prose, the system reaches 0.63--0.76~\bpb{} (7.9--9.5\%).
Crucially, an out-of-distribution evaluation on a document published after
SmolLM2's training cutoff confirms that these gains are not artifacts of
memorization: \nacrith{} achieves 0.723~\bpb{} on unseen text, outperforming
FineZip (same model) by 26\%.
All results use only 135M parameters and $\sim$500\,MB of GGUF weights.

Third, the hybrid NC06 format---to our knowledge, the first LLM-based
binary compressor---extends neural compression to arbitrary binary files, broadening the
applicability of this approach beyond the pure-text domain of all prior
work.

These results confirm that the compression--prediction equivalence
identified by Shannon~\citep{shannon1948} and formalized for LLMs by
Del\'etang et al.~\citep{deletang2024} is practically accessible today
on consumer hardware, with results well below classical Shannon entropy
bounds.

\medskip
\noindent\textbf{Code availability.}
\nacrith{} is open-source and available at
\url{https://github.com/robtacconelli/Nacrith-GPU}.

\bibliographystyle{plainnat}

\end{document}